\documentstyle[12pt]{article}
\begin{document}
\def\a{\alpha}
\def\b{\beta}
\def\ch{\chi}
\def\d{\delta}
\def\e{\epsilon}
\def\E{{\cal E}}
\def\f{\phi}
\def\g{\gamma}
\def\h{\eta}
\def\et{\tilde{\eta}}
\def\i{\iota}
\def\j{\psi}
\def\k{\kappa}
\def\l{\lambda}
\def\m{\mu}
\def\n{\nu}
\def\o{\omega}
\def\p{\pi}
\def\q{\theta}
\def\r{\rho}
\def\s{\sigma}
\def\t{\tau}
\def\u{\upsilon}
\def\x{\xi}
\def\z{\zeta}
\def\D{\Delta}
\def\F{\Phi}
\def\G{\Gamma}
\def\J{\Psi}
\def\L{\Lambda}
\def\O{\Omega}
\def\P{\Pi}
\def\S{\Sigma}
\def\U{\Upsilon}
\def\X{\Xi}
\def\T{\Theta}
\def\vf{\varphi}
\def\ve{\varepsilon}
\def\cC{{\cal P}}
\def\cD{{\cal Q}}

\def\pp {\partial }
\def\pb {\bar{\partial }}
\def\zb {\bar{z}}
\def\ET {\tilde{E }}
\def\gi {g^{-1}}
\def\be{\begin{equation}}
\def\ee{\end{equation}}
\def\ben{\begin{eqnarray}}
\def\een{\end{eqnarray}}

\hsize=16truecm
\addtolength{\topmargin}{-0.6in}
\addtolength{\textheight}{1.5in}
\vsize=26truecm
\hoffset=-.6in
\baselineskip=7 mm

\thispagestyle{empty}
\begin{flushright} SNUTP 98-096 \\ 
October 1998 \\
(Revised) \\
\end{flushright}
\begin{center}
 {\large\bf Exact BPS monopole solution in a self-dual background }
\vglue .5in
 Choonkyu Lee
 \vglue .2in
{\it
Department of Physics and Center for Theoretical Physics \\
Seoul National University \\
Seoul,151-742, Korea}
\vglue .2in

 Q-Han Park\footnote{ Electronic address; qpark@nms.kyunghee.ac.kr }
\vglue .2in
{\it
Department of Physics \\
Kyunghee University\\
Seoul, 130-701, Korea}
\vglue .2in
{\bf ABSTRACT}\\[.2in]
\end{center}
An exact one monopole solution in a uniform self-dual background field is 
obtained in the BPS limit of the $SU(2)$ Yang-Mills-Higgs theory by using 
the inverse scattering method.
\vglue .1in
\newpage
There has been much theoretical interest concerning magnetic monopole solutions 
in an $SU(2)$ Yang-Mills-Higgs theory after 't Hooft and Polyakov \cite{Pol} 
made the initial discovery of such structure in the seventies.
Especially, in the Bogomolny-Prasad-Sommerfend (BPS) limit \cite{Pra,Bog},
the ADHMN method \cite{Ati,Nah} can be used to construct exact
static multi-monopole solutions satisfying the first-order Bogomolny equations
\be
F_{ij}=-\e_{ijk}D_{k}\Phi ,
\label{BPS}
\ee
where $F_{ij}=\pp_{i}A_{j}-\pp_{j}A_{i}+i [A_{i},~A_{j}] \equiv \e_{ijk}B_{k} $ 
and $D_{k}\Phi =\pp_{k}\Phi +i[A_{k}, ~\Phi ]$ (with $A_{i}\equiv 
A_{i}^{a}\tau^{a} /2 , \Phi \equiv \Phi^{a}\tau^{a}/2 $). BPS monopoles 
refer to solutions of Eq. (\ref{BPS}), with the asymptotic fields approaching 
the Higgs vacuum (as is necessary for any finite-energy configuration). At 
large distances, they feature the field $B_{i}$ characteristic of a system of
localized magnetic monopoles and also the (gauge-invariant) magnitude of the 
Higgs field given as 
\be
|\Phi (\vec{r})| \approx v -{g \over 4\pi r} , 
\mbox{ for large } r
\label{Asy}
\ee
where $g=4\pi n~ (n=1,2,...)$ is the strength of the magnetic charge.
Note that studies of BPS monopoles are directly relevant in nonperturbative 
investigations of certain supersymmetric gauge theories. 

In this letter, we shall discuss a new solution of Eq. (\ref{BPS}) which becomes possible 
if we assume a more general asymptotic configuration than the Higgs vacuum.
As a particular solution of Eq. (\ref{BPS}), we have the uniform self-dual 
field described by (up to arbitrary gauge transformation)
\be
A_{i} = -{1\over 2}(\vec{r} \times \vec{B}_{0})_{i} \tau^{3} /2,  ~~~
\phi = -(v+\vec{B}_{0}\cdot \vec{r})\tau^{3}/2 .
\label{Ai}
\ee
If the magnetic field strength $\vec{B}_{0}$ were zero, this would reduce to 
the usual Higgs vacuum. In this work, we will look for a solution of 
Eq. (\ref{BPS}) which describes a (static) monopole in the asymptotic uniform 
field background of the form (\ref{Ai}) with $\vec{B}_{0} \ne 0$. 
For sufficiently weak $\vec{B}_{0}$, the corresponding, everywhere regular, 
solution was first discussed in Ref. \cite{Bak} (see Eqs. (3.35)-(3.37) of 
this article). From the latter, we know that the Higgs field in an appropriate 
gauge takes the form
\ben
\Phi (\vec{r}) &=& -\Big[ v (\mbox{coth} vr - {1\over vr}) +{1\over 2} \vec{B}_{0} 
\cdot \vec{r} (2 \mbox{coth} vr - {vr \over \mbox{sinh}^{2} vr})\Big] \hat{r} \cdot 
\vec{\tau } /2 \nonumber \\
&& - {1\over 2}{r \over \mbox{sinh} vr }\Big[ \vec{B}_{0}\cdot \vec{\tau}/2 
-(\vec{B}_{0}\cdot \hat{r}) \hat{r} \cdot \vec{\tau }/2 \Big] ,
\label{Weaksol}
\een                                                           
where $\hat{r}=\vec{r}/r$. 
(Note that, with $\vec{B}_{0} =0$, this reduces to the well-known 
Prasad-Sommerfield expression \cite{Pra}). This is a perturbative 
solution, i.e., valid only to the first order in $\vec{B}_{0}$, and 
therefore we still have no guarantee for the existence of the corresponding, 
globally well-defined, exact solution (with a finite background field 
$\vec{B}_{0}$) to the full nonlinear system  (\ref{BPS}). The full 
solution (see Eq. (\ref{fullsol})), 
which reduces to the perturbative result (\ref{Weaksol}) 
for small $\vec{B}_{0}$, will be found 
below with the help of the inverse scattering method. However, as we shall see,  
there arises some unusual feature when one tries to extend the solution to the 
whole 3-dimensional space.

As we make the choice $\vec{B}_{0} = B_{0} \hat{z}$ (with $B_{0} >0$), an 
obvious starting point for the solution, suggested by the symmetry consideration, 
will be the following cylindrical ansatz \cite{Man}:
\ben
A_{i}^{a} &=& -\hat{\vf}^{i}\Big[ {\eta_{2}(\rho ,z)-1 \over \rho }\hat{z}^{a}
+{\eta_{1}(\rho ,z)\over \rho }\hat{\rho }^{a}\Big] +\Big[ \hat{z}^{i} W_{1}(\rho ,z)
+\hat{\rho}^{i} W_{2} (\rho ,z)\Big] \hat{\vf }^{a} , \nonumber \\
\Phi^{a} &=& \phi_{1} (\rho , z)\hat{\rho }^{a} + \phi_{2} (\rho , z)\hat{z}^{a} ,
\label{Ans}
\een
where $(\rho , \vf , z)$ refer to cylindrical coordinates, and we have introduced 
normalized basis vectors (in ordinary 3-space and isospin space) $\hat{\rho }      
=(\cos {\vf }, \sin {\vf },0), \hat{\vf}=(-\sin {\vf }, \cos{\vf},0)$ and 
$\hat{z} =(0,0,1)$. Performing a judicious (singular) gauge transformation with 
Eq. (\ref{Ans}), it is also possible to write the ansatz in an alternative 
form \cite{For} (here note that $A_{i} \equiv A_{i}^{a } \tau^{a}/2$): 
\ben
A_{\rho } & \equiv & \cos \vf A_{1} + \sin \vf A_{2} = -W_{2} {\tau^{1} \over 2} 
={1\over 2}\pmatrix{0 & -W_{2} \cr -W_{2} & 0} , \nonumber \\
A_{\vf } & \equiv & -\sin {\vf }A_{1} +\cos {\vf}A_{2} =-{\eta_{1} \over \rho }
{\tau^{2} \over 2}-{\eta_{2} \over \rho}{\tau^{3} \over 2}= 
{1\over 2 \rho}\pmatrix{-\eta_{2} & i\eta_{1} \cr -i\eta_{1} & \eta_{2}},
\nonumber \\
A_{3} &= & -W_{1}{\tau^{1} \over 2}={1\over 2}\pmatrix{0 & -W_{1} \cr
-W_{1} & 0}, ~~~~ \Phi = {1\over 2}\pmatrix{\phi_{2} & -i\phi_{1} \cr
i\phi_{1} & -\phi_{2} }. 
\label{Ans2}
\een
Using either form, one finds from the Bogomolny equation in Eq. (\ref{BPS}) 
that the functions $\phi_{1}, \phi_{2}, \eta_{1}, \eta_{2} , W_{1}$ and 
$W_{2}$ should satisfy the coupled equations
\ben
\pp_{\rho }\phi_{1}-W_{2}\phi_{2}&=& -{1\over \rho }(\pp_{z}\eta_{1}-W_{1}\eta_{1}), 
~~~~\pp_{z}\phi_{1}-W_{1}\phi_{1}={1\over \rho}(\pp_{\rho}\eta_{1}-
W_{2}\eta_{2}), \nonumber \\
\pp_{\rho }\phi_{2}+W_{2}\phi_{1}&=& -{1\over \rho }(\pp_{z}\eta_{2}+W_{1}
\eta_{1}), ~~~~\pp_{z}\phi_{2}+W_{1}\phi_{1}={1\over \rho}(\pp_{\rho}\eta_{2}+
W_{2}\eta_{1}), \nonumber \\
\pp_{\rho}W_{1}-\pp_{z}W_{2}&=& -{1\over \rho}(\eta_{1}\phi_{2}-\eta_{2}\phi_{1}).
\label{Compeq}
\een    

By making a judicious gauge choice, it was shown in Refs. \cite{For,For2} that 
the solution to Eq. (\ref{Compeq}) can always be written as
\ben
\phi_{1}={\pp_{z} \psi \over f}, && ~~~ \phi_{2} =-{\pp_{z} f \over f} ,   ~~~~~~
\eta_{1} =-\rho {\pp_{\rho }\psi \over f} , \nonumber \\
\eta_{2}=\rho {\pp_{\rho}f \over f}, && ~~~ W_{1}=-\phi_{1}, ~~~~~ W_{2}={1\over \rho}\eta_{1}
\label{phis}
\een
with the two real functions $f=f(\rho ,z)$ and $\psi =\psi (\rho ,z)$ which must 
satisfy the Ernst equations \cite{Ern} (here, $\nabla^2 \equiv \pp_{z}^2 +\pp_{\rho}^2 
+{1\over \rho }\pp_{\rho })$
\be
f\nabla^2 \psi -2\nabla f \cdot \nabla \psi =0 , ~ 
f\nabla^2 f -|\nabla f|^2 + |\nabla \psi |^2 =0 .
\label{Ereq}
\ee
If we here define the real symmetric, $2\times 2$ unimodular matrix $g$ by 
\be
g={1\over f}\pmatrix{1 & \psi \cr \psi & \psi^2 + f^2 },
\ee
Eq. (\ref{Ereq}) can further be changed into the chiral equation (or Yang's 
equation \cite{Yan} for axially symmetric monopoles)
\be
\pp_{\rho}[\rho (\pp_{\rho}g)g^{-1}]+\pp_{z} [\rho (\pp_{z} g)g^{-1}]=0.
\label{Chiral}
\ee
Note that, for the Prasad-Sommerfield one-monopole solution, we have \cite{Loh}
\be
f={\rho \over F}, ~~~~~ \psi ={1\over F}(z \mbox{cosh} vz -r \mbox{sinh} vz ~
\mbox{coth} vr ) 
\label{Fterm}
\ee
where 
$
F \equiv r /  \mbox{sinh} vr  + r\mbox{cosh} vz~ \mbox{coth} vr 
-z \mbox{sinh} vz .$

In order to incorporate the effect of the background field on the result (\ref{Fterm}), 
we may use the inverse scattering method with the above chiral 
equation \cite{For2,For3}. It is based on the fact that Eq. (\ref{Chiral}) 
can be viewed as the compatibility conditions of the linear system
\ben
D_{1}\Psi &\equiv  & \Big( \pp_{z} - {2\l^2 \over \l^2 +\rho^2 }\pp_{\l } \Big) \Psi 
={\rho [\rho (\pp_{z}g )g^{-1} -\l (\pp_{\rho }g)g^{-1} ] \over 
\l^2 + \rho^2 }\Psi \nonumber \\
D_{2}\Psi &\equiv  & \Big( \pp_{\rho} + {2\l \rho \over \l^2 +\rho^2 }\pp_{\l } \Big) \Psi 
={\rho [\rho (\pp_{\rho }g )g^{-1} +\l (\pp_{z }g)g^{-1} ] \over 
\l^2 + \rho^2 }\Psi    
\label{Lineq}
\een
for a $2 \times 2$ matrix $\Psi =\Psi (\rho ,z; \l ).$ Now, for some initial 
solution $g=g_{0}(\rho ,z)$ of Eq. (\ref{Chiral}), suppose that we know a 
corresponding solution $\Psi_{0}(\rho , z ;\l )$ of Eq. (\ref{Lineq}), 
with the boundary condition $\Psi_{0}(\rho ,z;\l =0)=g_{0}(\rho ,z)$ 
satisfied. Then, the \underline{dressed} functions, $\Psi (\rho ,z;\l )
=\chi (\rho ,z;\l ) \Psi_{0}(\rho ,z;\l )$ and $g(\rho ,z)=\chi (\rho ,z;\l =0)
g_{0}(\rho ,z)$, give new solutions of Eqs. (\ref{Chiral}) and (\ref{Lineq}), 
provided that $\chi (\rho ,z;\l )$ satisfies
\ben
D_{1} \chi &=& { \rho [\rho (\pp_{z} g)g^{-1} -\l (\pp_{\rho }g)g^{-1}]
\over \l^2 + \rho^2 }\chi -\chi  ~
 { \rho [\rho (\pp_{z} g_{0})g_{0}^{-1} -\l (\pp_{\rho }g_{0})g_{0}^{-1}]
\over \l^2 + \rho^2 } , \nonumber \\
D_{2} \chi &=& { \rho [\rho (\pp_{\rho } g)g^{-1} +\l (\pp_{z }g)g^{-1}]
\over \l^2 + \rho^2 }\chi -\chi ~
 { \rho [\rho (\pp_{\rho } g_{0})g_{0}^{-1} +\l (\pp_{z }g_{0})g_{0}^{-1}]
\over \l^2 + \rho^2 } , 
\label{Lineq2}
\een
and also the condition (originating from the hermiticity of $g$ and $g_{0}$) 
\be
\chi (\rho , z;\l )=g(\rho ,z) [\chi (\rho ,z; -\rho^2 /\bar{\l } )]^{\dagger -1}
 g_{0} (\rho ,z)^{-1}.
\ee
 The function $\chi (\rho ,z;\l )$, needed in generating 
$N$-monopole solutions, may have only simple poles in the complex $\l $-plane 
(see Refs. \cite{For2,Loh}), viz.,
\be
\chi (\rho , z;\l ) = 1+ \sum^{N}_{k=1} {R_{k} (\rho ,z) \over \l -\m_{k} (\rho ,z)}
\label{Chi}
\ee
with the poles $\m_{k} (\rho , z)$ explicitly given by
\be
\m_{k} (\rho ,z) = w_{k} -z + \sqrt{(w_{k}-z)^2 + \rho^2 },
\ee
where $w_{k}$ are arbitrary constants. The residues $R_{k}(\rho ,z)$ are also 
found readily and then the resulting expression for $\chi (\rho ,z;\l )$ may be 
used to secure the following formula for the new solution $g=g_{\mbox{ph}}(\rho ,z)$ 
of Eq. (\ref{Chiral}):
\ben
g_{\mbox{ph}} &=& g/\sqrt{\mbox{det} g} , \nonumber \\
g_{ab}&=& (g_{0})_{ab} - \sum_{i=1}^{N} \sum_{j=1}^{N} (\m_{i}\bar{\m}_{j} )^{-1} 
(\G^{-1})_{ij}(g_{0})_{ac} \bar{m}_{c}^{j} m_{d}^{i} (g_{0})_{db} , (a,b =1,2 )
\label{deter}
\een
where $m_{b}^{k} = M_{c}^{k} [\Psi_{0} (\rho , z;\m_{k} )^{-1}]_{cb} $$ (M_{c}^{k}$ 
are constants) and $\G_{ij} = m^{i}_{a} (g_{0})_{ab}\bar{m}^{j}_{b}/(\rho^2 + 
\m_{i} \bar{\m }_{j} )$. 

For our problem, we may apply the above dressing method on the initial solutions 
which correspond to uniform self-dual fields. By a direct integration of the 
Ernst equation (\ref{Ereq}), we have a particular solution 
\be
g_{0} = \pmatrix{ 1/f_{0} & 0 \cr 0 & - f_{0}}; 
~~~~ f_{0} = \exp\Big[ vz +{B_{0} \over 2}(z^2 -{\rho^2 \over 2})\Big] \equiv 
\exp(vZ),  
\label{vZ}
\ee
and the corresponding  fields, if used in Eq. (\ref{Ans2}), yield precisely the 
uniform field configuration given in Eq. (\ref{Ai}). The minus sign in the component of 
$g_{0}$ is introduced in order to make det$g$ in Eq. (\ref{deter}) to be positive 
definite \cite{For2}. 
Given the matrix $g_{0} $ as in (\ref{vZ}), we may then solve the 
linear equations (\ref{Lineq}) for $\Psi_{0} = \pmatrix{\Psi_{0}^{1} & 0 \cr 0 & \Psi_{0}^{2}}$. 
All together, we have here four equations for $\Psi_{0}^{1}$ and $\Psi_{0}^{2}$, which may be 
integrated by noticing that two equations from the four in fact imply
\ben    
&& [\pp_{z} + (v + B_{0} z)+ {\l \over \rho } \pp_{\rho }] \Psi_{0}^{1} = 0,
\nonumber \\
&& [\pp_{z} -(v + B_{0} z) + {\l \over \rho } \pp_{\rho }] \Psi_{0}^{2} = 0.
\een
For a solution $\Psi_{0}(\rho ,z;\l )$ which satisfies the boundary condition 
$\Psi_{0} (\rho ,z;\l =0) =g_{0} (\rho ,z)$, we have found through this analysis 
the following expression:
\be
\Psi^{1}_{0} (\rho ,z ;\l )
= {1 \over \Psi_{0}^{2} (\rho , z;\l ) }
= \exp \Big[ -v(z+ {\l \over 2})-{B_{0} \over 2} (z^2 -
{1\over 2}\rho^2 + \l z + {1\over 4}\l^2 )\Big]^{-1} \equiv K .
\ee
Then, for the one monopole case (i.e., $N=1$ in Eq. (\ref{Chi})), 
the dressing method yields the $2 \times 2$ matrix $g(\rho ,z)$ with 
\ben
g_{11} &=& - { \m^2 f_{0}^2 K ^4 M_{2}^2 + \r^{2} M_{1}^{2} 
                \over \m^{2}f_{0} ( M_{1}^{2} -
                K^{4}M_{2}^{2}f_{0}^{2} )}
            \nonumber \\
g_{12} &=& g_{21} = {(\r^{2} + \m^{2})f_{0} K^2 M_{1}M_{2} \over
                \m^{2}(M_{1}^{2} - K^{4}M_{2}^{2}f_{0}^{2})}
                 \nonumber \\
g_{22} &=& - { \m^2 f_{0} M_{1}^{2} + \r^{2} K^{4}M_{2}^{2}f_{0}^{3} \over
            \m^{2}(M_{1}^{2} - K^{4}M_{2}^{2}f_{0}^{2} )},
\een
where
$\m = -z + r, ~ r\equiv \sqrt{z^2 + \r^2 }$, and $f_{0}$ is given in Eq. (\ref{vZ}).
Finally, a new solution can be constructed directly from Eq. (\ref{deter}).
However, in order to compare with previously known results in the limiting case,
we make a gauge transformation of $g_{ph}$ through
$g_{ph}\rightarrow h g_{ph}h^{-1}$
where $h = \pmatrix{1/\sqrt{2} & 1/\sqrt{2} \cr -1/\sqrt{2} &
1/\sqrt{2}}$. Note that this is indeed a gauge transformation which leaves the 
chiral equation (\ref{Chiral}) covariant. 
This gives rise to the identification:
\ben
1/f &=& {\m \over 2 \r }(g_{11} - g_{21} -g_{12} + g_{22})
\nonumber \\
\psi &=& {\m f \over 2\r }(g_{11}-g_{22}) .
\een
Explicit evaluation then gives the expressions
\ben
f &=& {\rho \over \tilde{F}} , ~  
\tilde{F} \equiv {r \over \mbox{sinh} vR } + r \mbox{cosh} vZ \mbox{cosh} vR -
z \mbox{sinh} vZ , \nonumber \\
\psi &=& {1 \over \tilde{F}}(z \mbox{cosh} vZ - r \mbox{sinh} vZ ~\mbox{coth} vR ),
\label{psiF}
\een
where $R \equiv r(1+{B_{0} z\over 2 v})$ and $Z\equiv z +{B_{0} \over 2v}(z^2 
-{1\over 2}\rho^2 )$ (see Eq. (\ref{vZ})). 

Note that, with $B_{0} =0$ (i.e., in 
the zero background field limit), our expressions (\ref{psiF}) 
reduce to the known results (\ref{Fterm}); in this sense, Eq. (\ref{psiF}) provides a 
deformation of the Prasad-Sommerfield solution by allowing the background magnetic field. 
If the functions $(\phi_{1}, \phi_{2}, \eta_{1}, \eta_{2}, W_{1}, W_{2})$, 
calculated using Eqs. (\ref{phis}) and (\ref{psiF}), are inserted into Eq. (\ref{Ans}), 
we have an exact solution to the Bogomolny equations (\ref{BPS})
which are regular at $r=0$ and also on the $z$-axis. Explicitly, for the Higgs field, we find
\ben
\Phi^{a} (\vec{r}) &=& -v \{ (1 + {B_{0} \over v}z ) \mbox{coth} vR -{1\over vr} \} 
[ \hat {\rho}^{a} \cos \L (\rho , z) + \hat{z} \sin \L (\rho ,z )] \nonumber \\
&& + {B_{0} \rho \over 2 \mbox{sinh} vR } [\hat{\rho }^{a} \sin \L (\rho ,z )
-\hat {z} \cos \L (\rho ,z) ]  
\label{Oursol}
\een
with the function $\L (\rho ,z)$ defined through
\be
\tan \L (\rho ,z) = {z (1+ \mbox{cosh} vZ ~ \mbox{cosh} vR ) -r \mbox{sinh} vZ ~ 
\mbox{sinh} vR \over \rho (\mbox{cosh} vZ + \mbox{cosh} vR )}.
\ee
This leads to the gauge-invariant Higgs field magnitude
\be
|\Phi (\vec{r})|^2 = v^2 \Big[ (1+ {B_{0}z \over v})\mbox{coth} vR 
-{1\over vr }\Big] ^{2} + {B_{0}^{2} \rho^{2} \over 4 \mbox{sinh}^{2} vR }.
\label{Phisq}
\ee    
The small $B_{0}$-limit of this expression can easily be shown to coincide 
with the gauge-invariant magnitude obtained using the perturbative 
solution (\ref{Weaksol}); up to gauge transformation, the solution we have above 
is what we were after. Also, the appearance of the function $Z(z, \rho )$ above can be 
ascribed to a gauge artifact. After performing an appropriate (complicated) gauge 
transformation with the above solution, we have succeeded in casting our full 
solution, including $A_{i}^{a} (\vec{r})$, into the form (with $R \equiv r[1+{1\over 2v}
\vec{B}_{0}\cdot \vec{r} ]$)
\ben
\Phi^{a} (\vec{r} ) &=& -\hat{r}^{a} v \Big[ (1+{1\over v} \vec{B}_{0}\cdot \vec{r} )
\mbox{coth }vR - {1\over vr }\Big] - {r \over 2 \mbox{sinh }vR }\Big[ (\vec{B}_{0})^{a} 
-\vec{B}_{0}\cdot \hat{r}~ \hat{r}^{a} \Big] , \nonumber \\
A_{i}^{a}(\vec{r}) &=& -\e_{aij} {\hat{r}_{j} \over r} \Big[ 1 - 
(1+{1\over v}\vec{B}_{0}\cdot \vec{r} ){vr \over \mbox{sinh }vR } \Big] \nonumber \\
&& + \e_{aij} \Big[ (\vec{B}_{0})_{j}-\vec{B}_{0}\cdot \hat{r}~ \hat{r}_{j} \Big] 
{r \over 2 \mbox{sinh }vR } + {r \over 2} \Big( {1-\mbox{cosh }vR \over \mbox{sinh }vR }
\Big) \hat{r}^{a} \e_{ilm} \hat{r}_{l} (\vec{B}_{0})_{m} .
\label{fullsol}
\een
[We have verified that Eq. (\ref{Compeq}), and hence Eq. (\ref{BPS}), is satisfied 
by this expression].   
 From Eq. (\ref{Phisq}) we note that the Higgs zero or the 
monopole center, orginally at the origin for $B_{0}=0$, gets displaced along the 
$z$-axis for nonzero $B_{0}$. Evidently,  $|\Phi |^2$ 
may have zeros only along the line $\r =0$.)  
In fact, a detailed analysis shows that zeros of the 
Higgs field occurs in a rather nontrivial way depending on the strength of the 
background field $B_{0}$ (See below). 
Nevertheless, at large distances where $vR >>1$, we find
\be
|\Phi (\vec{r})| \approx v + B_{0} z - {1\over r},   ~~~(vR >>1 )
\label{Higapp}
\ee  
which is the expected behavior if an $n=1$ monopole is situated near the origin 
in the presence of the background field $\vec{B}_{0} = B_{0} \hat{z}$.
But, at points on the plane $z = -2v/B_{0}$ (which is, for small $B_{0}$, on the 
far left of our monopole), $R=r(1+{B_{0} \over 2v}z) \rightarrow 0$ and 
$|\Phi (\vec{r})|$ in Eq. (\ref{Phisq}) diverges, therefore, our solution possesses a  
surface singularity.                                        

It turns out that Higgs field has a single zero at $z=0$ 
only for the vanishing background case, $B_{0} =0$. For $B_{0} \ne 0$, 
Higgs field has a couple of zeros when $B_{0}$ is small 
and has no zero at all when $B_{0}$ exceeds a critical value, $B_{0} > 
B_{0}^{c} \approx 0.3 v^2 $. This makes it difficult to address notions such as the 
monopole center or the monopole number in terms of Higgs zeros as in the case of 
the vanishing $B_{0}$.
In order to help understand the situation better, it would be useful 
to consider the background self-dual solution itself as given in Eq. (\ref{Ai}). 
It has the plane $z = -v/B_{0}$ as the zero of the background Higgs field. 
Note that, if there exists some extended region where the Higgs field becomes very small, 
the topological character usually related to a magnetic monopole gets rather murky.
Our solution has a monopole deep in the right half-space $z >-v/B_{0}$ and 
shows a plausible behavior in the very right half. 
On the other hand, the plane $z = -v/B_{0}$ as the zeros of the background Higgs field 
has disappeared. In our solution (\ref{Oursol}) or (\ref{fullsol}), 
we have instead an isolated zero
near the point $\rho =0, ~~z=-v/B_{0}$ immersed in the region of small, but non-zero, 
Higgs field and there is no distinctive long-range tail associated with this zero.
Even in the other half-space 
where the Higgs field was aligned in the opposite direction, $|\Phi (\vec{r})|$ is 
well approximated by (\ref{Higapp}) if $z$ is not too close to $-2v/B_{0}$. 
 However, the divergence of $|\Phi (\vec{r})|$ encountered at $z=-2v/B_{0}$ is
nontrivial; above all, it is not an gauge artifact. Thus, our monopole solution 
cannot be extended beyond this singular plane. If one is concerend with only restricted 
physical problems (as in Ref. [6]), this ill-behavior of our solution in the `wrong' 
Higgs vacuum region might not be taken too seriously. But our opinion is that this 
singularity issue deserves further investigation in the future.

A couple of comments are in order. We note that the well-known trick \cite{Wei} 
may be used on our solution to obtain the corresponding dyon solution which solves the 
generalized Bogomolny equations \cite{Col}
\be
B_{i} = -\cos {\b } ~ D_{i} \Phi , ~~~~ E_{i} = - \sin {\b } ~ D_{i} \Phi 
\ee
in the background of a uniform magnetic and electric field. Also, our approach 
can be applied to the problem of finding exact instanton solutions in 
nonvanishing background fields as well. In this regard, it would be interesting to extend the 
ADHM construction \cite{Ati} and the Nahm equation \cite{Nah} in the presence of background fields.

\vskip .5in

 {\bf ACKNOWLEDGEMENT}
\vglue .2in
Useful discussions with D. Bak, K. Lee and H. Min are acknowledged. This work was 
supported in part by the Korea Science and Engineering Foundation (through 
the SRC program of the SNU-CTP and 97-07-02-02-01-3 ) and the Basic Science 
Research program under BSRI-97-2418 and BSRI-97-2442.
\vglue .2in

\end{document}